\documentclass[twocolumn,showpacs,preprintnumbers,amsmath,amssymb,aps,prd]{revtex4}

\newcommand{\Lie}[0]{{\cal L}\, }
\newcommand{\grad}[0]{\nabla\!}
\newcommand{\R}{{\mathcal{R}}}
\def\th{{\widehat{\tau}}}
\def\rh{{\widehat{r}}}
\def\twoR{{\widetilde{\R}}}
\def\ls{{(\ell)}}
\def\ns{{(n)}}
\def\l{{\ell}}
\def\fie{\varphi}
\def\q{{\widetilde{q}}}
\def\K{{\widetilde{K}}}

\begin{document}

\title{Dynamical Horizons: Energy, Angular Momentum, Fluxes and Balance Laws}
\date{\today}
    \author{Abhay Ashtekar}
    \email{ashtekar@gravity.psu.edu}
    \author{Badri Krishnan}
    \thanks{Present address: Max-Planck-Institut f\"ur Gravitationsphysik,
    Albert-Einstein-Institut, Am M\"uhlenberg 1, D-14476 Golm, Germany}
    \email{badkri@aei-potsdam.mpg.de}
    \affiliation{Center for Gravitational Physics and Geometry
    and Center for Gravitational Wave Physics, Department of
    Physics, Penn State University, University Park, PA 16802,
    USA}

\begin{abstract}

Dynamical horizons are considered in full, non-linear general
relativity. Expressions of fluxes of energy and angular momentum
carried by gravitational waves across these horizons are obtained.
Fluxes are local, the energy flux is positive and change in the
horizon area is related to these fluxes. The flux formulae also
give rise to balance laws analogous to the ones obtained by Bondi
and Sachs at null infinity and provide generalizations of the
first and second laws of black hole mechanics.

\end{abstract}


\maketitle

Black holes which are themselves in equilibrium but in possibly
time-dependent space-times can be modelled by \emph{isolated
horizons} \cite{letter}. Over the past three years, properties of
isolated horizons were studied in detail. In particular, the
framework enabled one to extend the laws of black hole mechanics
\cite{afk,abl2} and has been used to extract physics from initial
data of widely separated black holes \cite{bk} and from numerical
simulations of the final phases of black hole collisions
\cite{letter,dkss}. The purpose of this letter is to outline a
generalization of these ideas to fully dynamical situations in
which matter and gravitational radiation can fall into black
holes.

Our definition of a dynamical horizon is geared to practical
applications in astrophysical situations, particularly those
considered in numerical relativity.

\section{Definition and notation}
\label{sec:defn}

\noindent \textbf{Definition:} A smooth, three-dimensional,
space-like sub-manifold $H$ in a space-time is said to be a
\emph{dynamical horizon} if it is foliated by a preferred family
of 2-spheres such that, on each leaf $S$, the expansion
$\theta_{(\ell)}$ of a null normal $\ell^a$ vanishes and the
expansion $\theta_{(n)}$ of the other null normal $n^a$ is
strictly negative.

Thus, a dynamical horizon $H$ is a 3-manifold which is foliated by
marginally trapped 2-spheres. As shown below, the area of these
2-spheres necessarily increases. An example is provided by
continuous segments of world tubes of apparent horizons that
naturally arise in numerical evolutions of black holes. In
contrast to event horizons, dynamical horizons can be located
quasi-locally; knowledge of the full space-time is not required.
The condition that $H$ be space-like is implied by a stronger but
physically reasonable restriction that the derivative of
$\theta_{(\ell)}$ along $n^a$ be negative \cite{sh}. Finally, the
requirement that the leaves be topologically $S^2$ can be replaced
by the weaker condition that they be compact. One can show that
the topology of $S$ is necessarily $S^2$ if the flux of matter or
gravitational energy across $H$ is non-zero. If these fluxes were
to vanish identically, $H$ would become isolated and replaced by a
null, non-expanding horizon \cite{afk}.

Dynamical horizons are closely related to Hayward's trapping
horizons \cite{sh}. However, while the definition of trapping
horizons imposes a condition on the derivative of
$\theta_{(\ell)}$ off $H$, our conditions refer only to geometric
quantities which are intrinsically defined on $H$. But in cases of
physical interest, the additional condition would be satisfied and
dynamical horizons will be future, outer trapping horizons.
Nonetheless, our analysis and results differ considerably from
those of Hayward's. While his framework is based on a 2+2
decomposition, ours will be based on the ADM 3+1 decomposition.
Our discussion includes angular momentum, our flux formulae are
new and our generalization of black hole mechanics is different.
Our analysis is geared to providing tools to extract physics and
perform checks on numerical simulations of dynamical black holes.
Therefore we will restrict ourselves to dynamical horizons with
zero charge.

Let us begin by fixing notation. Let $\th^a$ be the unit time-like
normal to $H$ and  denote by $\grad_a$ the space-time derivative
operator. The metric and extrinsic curvature of $H$ are denoted by
$q_{ab}$ and $K_{ab}:={q_a}^c{q_b}^d\grad_c\th_d$ respectively;
$D_a$ is the derivative operator on $H$ compatible with $q_{ab}$
and $\R_{ab}$ its  Ricci tensor. Leaves of the preferred foliation
of $H$ will be called \emph{cross-sections} of $H$. The unit
space-like vector orthogonal to $S$ and tangent to $H$ is denoted
by $\rh^{\,a}$. Quantities intrinsic to $S$ will be generally
written with a tilde. Thus, the two-metric on $S$ is $\q_{ab}$,
the extrinsic curvature of $S\subset H$ is
$\K_{ab}:=\widetilde{q}_a^{\,\,\,\,c}\widetilde{q}_b^{\,\,\,\,d}
D_c\rh_d$, the derivative operator on $(S, \q_{ab})$  is
$\widetilde{D}_a$ and its Ricci tensor is $\twoR_{ab}$. Finally,
we will fix the rescaling freedom in the choice of null normals
via $\l^a:=\th^{\,a}+\rh^{\,a}$ and $n^a:=\th^{\,a}-\rh^{\,a}$.

We first note an immediate consequence of the definition. Since
$\theta_\ls =0$ and $\theta_\ns <0$, it follows that $\K >0$.
Hence the area $a_S$ of $S$ increases monotonically along $\rh^{\,
a}$. Thus the second law of black hole mechanics holds on $H$. We
will obtain an explicit expression for the change of area in part
\ref{sec:energy}.

Our main analysis is based on the fact that, since $H$ is a
space-like surface, the Cauchy data $(q_{ab},K_{ab})$ on $H$ must
satisfy the usual scalar and vector constraints
\begin{eqnarray} H_S &:=& \R + K^2 - K^{ab}K_{ab}
          = 16\pi G T_{ab}\th^{\,a}\th^{\,b}
            \label{eq:hamconstr}\\
    H_V^a &:=& D_b\left(K^{ab} - Kq^{ab}\right)
    = 8\pi G T^{bc}\th_{\, c}{q^a}_b
    \label{eq:momconstr} \, .\end{eqnarray}
We will often focus our attention on a portion $\Delta H \subset
H$ which is bounded by two cross-sections $S_1$ and $S_2$.

\section{Angular momentum}
\label{sec:angmom}

The angular momentum analysis is rather straight forward and is,
in fact, applicable to an arbitrary space-like hypersurface.  Fix
\emph{any} vector field $\fie^a$ on $H$ which is tangential to the
cross-sections of $H$. Contract $\fie^a$ with both sides of eqn.
(\ref{eq:momconstr}). Integrate the resulting equation over the
region $\Delta H\subset H$, perform an integration by parts and
use the identity $\Lie_\fie q_{ab} = 2D_{(a}\fie_{b)}$ to obtain
\begin{eqnarray} && \label{eq:balanceJ} \frac{1}{8\pi G}\oint_{S_2}K_{ab}
\fie^a\rh^{\,b} \, d^2V - \frac{1}{8\pi G}
\oint_{S_1}K_{ab}\fie^a\rh^{\,b} \, d^2V  \nonumber \\
&& = \int_{\Delta H} \left( T_{ab}\th^{\,a}\fie^b + \frac{1}{16\pi
G} P^{ab}\Lie_\fie q_{ab}\right)\, d^3V\end{eqnarray}
where $P^{ab}:=K^{ab}-Kq^{ab}$. It is natural to identify the
surface integrals with the generalized angular momentum
$J^{(\fie)}$ associated with those surfaces and set:
\begin{equation} \label{eq:jdynamic}J_S^{(\fie)} = -\frac{1}{8\pi G} \oint_{S}
K_{ab} \fie^a\rh^{\,b} \, d^2V \end{equation}
where we have chosen the overall sign to ensure compatibility with
conventions normally used in the asymptotically flat context. The
term `generalized' emphasizes the fact that the vector field
$\fie^a$ need not be an axial Killing field even on $S$; it only
has to be tangential to our cross-sections.

The flux of this angular momentum due to matter fields and
gravitational waves are respectively
\begin{eqnarray} \mathcal{J}^{(\fie)}_\textrm{m} &=& -\int_{\Delta H}
 T_{ab}\th^{\,a}\fie^b\, d^3V \, ,\\
 \mathcal{J}^{(\fie)}_{\textrm{g}} &=& -\frac{1}{16\pi G}
 \int_{\Delta H} P^{ab}\Lie_\fie q_{ab}\, d^3V \, , \end{eqnarray}
and we get the balance equation
\begin{equation}  J_2^{(\fie)} - J_1^{(\fie)} =
 \mathcal{J}^{(\fie)}_\textrm{m}+
\mathcal{J}^{(\fie)}_{\textrm{g}}\, . \end{equation}
As expected, if $\fie^a$ is a Killing vector of the three-metric
$q_{ab}$, then the gravitational flux vanishes:
$\mathcal{J}^{(\fie)}_{\textrm{g}} = 0$.  It is convenient to
introduce the \emph{angular momentum current}
${j}^{\fie}:=-K_{ab}\fie^a\rh^{\,b}$ so that (\ref{eq:jdynamic})
becomes $J_S^{(\fie)}=(8\pi G)^{-1}\oint_S {j}^{\fie}\, d^2V$.

\section{Energy fluxes and area balance}
\label{sec:energy}

As is usual in general relativity, the notion of energy is tied to
a choice of a vector field.  Here, we will consider vector fields
$\xi^a = N\l^a$ where the lapse $N$ is constructed as follows. Let
$r$ be a radial coordinate on $H$ defined such that the
cross-sections of $H$ are level surfaces of $r$. Then
$\rh_a\propto D_ar$.  It turns out that in order to get the
balance law for energy, we must tie our lapse functions $N$ to
radial coordinates such that $D_ar = N_r\rh_a$. (Since $\xi^a =
N\l^a$, as usual the term `lapse' refers to space-time evolution;
not to `evolution' along $\hat{r}^a$.) Thus each $r$ determines a
\emph{permissible} lapse function $N_r$. If we use a different
radial coordinate $r^\prime$, then the lapse is rescaled according
to the relation
\begin{equation} N_{r^\prime} = N_r \frac{dr^\prime}{dr} \, .
\end{equation}
Thus, although the lapse itself will in general be a function of
all three coordinates on $H$, the relative factor between any two
permissible lapses can be a function only of $r$. We denote the
resulting \emph{permissible} vector fields by $\xi^a_{(r)}:=
N_r\l^a$. Recall that, on an isolated horizon, physical fields are
time independent and null normals can be rescaled by a positive
constant \cite{afk}. Now the horizon fields are `dynamical', i.e.,
$r$-dependent, and the rescaling freedom is by a positive function
of $r$.

We are interested in calculating the flux of energy associated
with $\xi_{(r)}^a$ for any radial coordinate $r$. Denote the flux
of matter energy across $\Delta H$ by $\mathcal{F}^{(r)}_m:=
\int_{\Delta H} T_{ab}\th^{\,a}\xi_{(r)}^b d^3V$. By taking the
appropriate combination of  (\ref{eq:hamconstr}) and
(\ref{eq:momconstr}) we obtain
\begin{equation} \label{eq:fluxTr} \mathcal{F}^{(r)}_m=
\frac{1}{16\pi G} \int_{\Delta H}\, N_r \left\{H_S + 2\rh_a H_V^a
\right\}\, d^3V\, .
\end{equation}
Since $H$ is foliated by two-spheres, we can perform a $2+1$ split
of the various quantities on $H$.  Using the Gauss Codazzi
relation we rewrite $\R$ in terms of quantities on $S$:
\begin{equation} \label{eq:threeR} \R = \twoR + \K^2 -\K_{ab}\K^{ab} +
2D_a\alpha^a \end{equation}
where $\alpha^a = \rh^{\,b}D_b\rh^{\,a} - \rh^{\,a}D_b\rh^{\,b}$.
Next, the fact that the expansion $\theta_\ls$ of $\l^a$ vanishes
leads to the relation
\begin{equation} \label{eq:expansion0} K + \K = K_{ab}\rh^{\,a}\rh^{\,b} \,
.\end{equation}
Using  (\ref{eq:threeR}) and (\ref{eq:expansion0}) in
eqn.~(\ref{eq:fluxTr}) and simplifying, we obtain the result
\begin{eqnarray} \label{eq:NR} \int_{\Delta H} N_r \twoR\,d^3V &=& 16\pi G
\int_{\Delta H} T_{ab}\th^{\,a}\xi_{(r)}^b\,d^3V \nonumber \\ &+&
\int_{\Delta H} N_r\left\{ |\sigma|^2 + 2|\zeta|^2\right\}\,d^3V
\end{eqnarray}
where $|\sigma|^2=\sigma_{ab}\sigma^{ab}$ with $\sigma_{ab}
:=\widetilde{q}_a^{\,\,\,c}\widetilde{q}_b^{\,\,\,d}\grad_c\ell_d
-\frac{1}{2}\widetilde{q}_{ab}\widetilde{q}^{\,\,cd}\grad_c\ell_d$,
the shear of $\l^a$, and $|\zeta|^2= \zeta^a\zeta_a$ with
$\zeta^a:=\widetilde{q}^{\,ab}\rh^{\,c}\grad_c\l_b$. Both
$\sigma_{ab}$ and $\zeta^a$ are tensors intrinsic to $S$. To
simplify the left side of this equation, note that the volume
element $d^3V$ on $H$ can be written as $d^3V = N_r^{-1}dr\,d^2V$
where $d^2V$ is the area element on $S$. Using the Gauss-Bonnet
theorem, the integral of $N_r\twoR$ can then be written as
\begin{equation} \int_{\Delta H} N_r\twoR \,d^3V = \int_{r_1}^{r_2} dr
\left(\oint_S \twoR\,d^2V\right) = 8\pi(r_2-r_1) \,
.\end{equation}
Substituting this result in eqn.~(\ref{eq:NR}) we finally obtain
\begin{eqnarray} \label{eq:balancel} \left(\frac{r_2}{2G}-
\frac{r_1}{2G}\right)
&&= \int_{\Delta H} T_{ab}\th^{\,a}\xi_{(r)}^b\,d^3V \nonumber \\
+&& \!\!\!\!\frac{1}{16\pi G} \int_{\Delta H} N_r\left\{
|\sigma|^2 + 2|\zeta|^2\right\} \,d^3V . \end{eqnarray}
This is the key result we were looking for. Let us now interpret
the various terms appearing in this equation. The first integral
on the right side of this equation is the flux
$\mathcal{F}^{(r)}_m$ of matter energy associated with the vector
field $\xi_{(r)}^a$.  Since $\xi_{(r)}^a$ is null and $\th$
time-like, if $T_{ab}$ satisfies, say, the dominant energy
condition, this quantity is guaranteed to be non-negative. It is
natural to interpret the second term as the flux
$\mathcal{F}^{(r)}_g$ of $\xi_{(r)}^a$-energy in the gravitational
radiation:
\begin{equation} \label{eq:ab}
 \mathcal{F}^{(r)}_g := \frac{1}{16\pi G}
 \int_{\Delta H} N_r\left\{ |\sigma|^2 + 2|\zeta|^2\right\}\,d^3V\, .
\end{equation}
This expression shares four desirable features with the
Bondi-Sachs energy flux at null infinity. First, it does not refer
to any coordinates or tetrads; it refers only to the given
dynamical horizon $H$ and the evolution vector field
$\xi_{(r)}^a$. Second, the energy flux is manifestly non-negative.
Third, all fields used in it are local; we did not have to
perform, e.g., a radial integration to define any of them.
Finally, the expression vanishes in the spherically symmetric
case: if the Cauchy data $(q_{ab},K_{ab})$ and the foliation on
$H$ is spherically symmetric, $\sigma_{ab}=0$ and $\zeta^a=0$.

To conclude this section, let us choose for our radial coordinate
the area radius $R:=\sqrt{a/4\pi}$. Then,
\begin{equation} \label{eq:areaincr} \frac{R_2}{2G}-\frac{R_1}{2G} =
\mathcal{F}^{(R)}_m + \mathcal{F}^{(R)}_g \end{equation}
Thus, as promised in part \ref{sec:defn}, we have obtained an
explicit formula relating the change in the area of the horizon to
fluxes of matter and gravitational $\xi_{(R)}$-energy.

\section{Mass and the first law}
\label{sec:mass}

Let us now combine the results of parts \ref{sec:angmom} and
\ref{sec:energy} to obtain the physical process version of the
first law for $H$ and a mass formula for an arbitrary
cross-section of $H$.

Denote by $E^{\xi_{(R)}}$ the $\xi_{(R)}$-energy of cross-sections
$S$ of $H$. While we do not yet have the explicit expression for
it, we can assume that, because of the influx of matter and
gravitational energy, $E^{\xi_{(R)}}$ will change by an amount
$\Delta\, E^{\xi_{(R)}} = \mathcal{F}^{(R)}_m+
\mathcal{F}^{(R)}_g$ as we move from one cross section to another.
Therefore, if we define \emph{effective surface gravity} $k_R$
associated with $\xi^a_{(R)}$ as $k_R:=1/2R$, the infinitesimal
form of (\ref{eq:areaincr}) implies $({k_R}/{8\pi G}) da = d
E^{\xi_{(R)}}$. For a general choice of the radial coordinate $r$,
(\ref{eq:balancel}) yields a generalized first law:
\begin{equation} \label{eq:sphfirstlaw} \frac{k_r}{8\pi G} da
= d E^{\xi_{(r)}}
\end{equation}
where the effective surface gravity $k_r$ of $\xi^a_{(R)}$ is
given by
\begin{equation} k_r = \frac{dr}{dR}k_R \qquad \textrm{where}
 \qquad \xi^a_{(r)} = \frac{dr}{dR}\xi^a_{(R)} \, .\end{equation}
This rescaling freedom in surface gravity is analogous to the
rescaling freedom which exists for Killing horizons, or more
generally, isolated horizons. The new feature in the present case
is that we have the freedom to rescale the surface gravity (and
$\l^a$) by a positive function of the radius instead of just by a
constant. This is just what one would expect in a dynamical
situation. Finally, note that the differentials appearing in
(\ref{eq:sphfirstlaw}) are actual variations along the dynamical
horizon due to an infinitesimal change in $r$ and are not
variations in phase space as in some of the formulations
\cite{rw,afk,abl2} of the first law.

To include rotation, pick a vector field $\fie^a$ on $H$ such that
$\fie^a$ is tangent to the cross-sections of $H$, has closed
orbits and has affine length $2\pi$.( At this point, $\fie^a$ need
not be a Killing vector of $q_{ab}$.) Consider time evolution
vector fields $t^a$ which are of the form
$t^a=N_r\l^a-\Omega\fie^a$ where $N_r$ is a permissible lapse
associated with a radial coordinate $r$ and $\Omega$ an arbitrary
function of $r$. Evaluate the quantity $\int_{\Delta H}
T_{ab}\th^{\,a}t^b\, d^3V$ using (\ref{eq:balanceJ}) and
(\ref{eq:balancel}):
\begin{eqnarray}  &&\frac{r_2-r_1}{2G} + \frac{1}{8\pi G} \biggl\{
\oint_{S_2}\Omega j^\fie\,d^2V -\oint_{S_1} \Omega j^\fie\,d^2V
\biggr. \nonumber \\ &&- \biggl.\int_{\Omega_1}^{\Omega_2} d\Omega
\oint_S j^\fie\,d^2V
\biggr\} = \int_{\Delta H} T_{ab}\th^{\,a}t^b \,d^3V \nonumber \\
&& + \frac{1}{16\pi G}\int_{\Delta H} N_r \left(|\sigma|^2 +
2|\zeta|^2\right)\,d^3V \nonumber \\ &&- \frac{1}{16\pi G}
\int_{\Delta H}\Omega P^{ab}\Lie_\fie q_{ab} \, d^3V \,
.\label{eq:integrated1law} \end{eqnarray}
Again, if we denote by $E^t$ the $t$-energy associated with
cross-sections $S$ of $H$, the right side of
(\ref{eq:integrated1law}) can be interpreted as $\Delta E^t$.  If
we now restrict ourselves to infinitesimal $\Delta H$, the three
terms in the curly brackets combine to give $d(\Omega J) - J
d\Omega$ and we obtain
\begin{equation} \label{eq:genfirstlaw}  \frac{dr}{2G}+
\Omega dJ = \frac{k_r}{8\pi G}da + \Omega dJ\,= \, dE^t .
\end{equation}
This equation is our generalization of the first law for dynamical
horizons. Since the differentials in this equation are variations
along $H$, this can be viewed as a `physical process version of
the first law'. Note that for each allowed choice of lapse $N_r$,
angular velocity $\Omega(r)$ and vector field $\fie^a$ on $H$, we
obtain a permissible time evolution vector field
$t^a=N_r\l^a-\Omega\fie^a$ and a corresponding first law. This
situation is very similar to what happens in the isolated horizon
framework where we obtain a first law for each permissible time
translation on the horizon. Again, the generalization from that
time independent situation consists of allowing the lapse and the
angular velocity to become r-dependent, i.e., `dynamical'.

For every allowed choice of $(N_r,\Omega(r),\fie^a)$, we can
integrate eqn.~(\ref{eq:genfirstlaw}) on $H$ to obtain a formula
for $E^t$ on any cross section but, in general, the result may not
be expressible just in terms of geometric quantities defined
locally on that cross-section. However, in some physically
interesting cases, the expression \emph{is} local. For example, In
the case of spherical symmetry, it is natural to choose $\Omega=0$
and $R$ as the radial coordinate in which case we obtain
$E^t=R/2G$. This is just the irreducible (or Hawking) mass of the
cross-section. Even in this simple case, (\ref{eq:integrated1law})
provides a useful balance law, with clear-cut interpretation.
Physically, perhaps the most interesting case is the one in which
$q_{ab}$ is only axi-symmetric with $\fie^a$ as its axial Killing
vector. In this case we can naturally apply, at each cross-section
$S$ of $H$, the strategy used in the isolated horizon framework to
select a preferred $t^a$: Calculate the angular momentum $J$
defined by the axial Killing field $\fie$, choose the radial
coordinate $r$ (or equivalently, the lapse $N_r$) such that
\begin{equation} \label{eq:kerrkappa} k_r = k_o(R) :=
\frac{R^4-4G^2J^2}{2R^3\sqrt{R^4+4G^2J^2}} \end{equation}
and choose $\Omega$ such that
\begin{equation} \Omega= \Omega_o(R) := \frac{2GJ}{R\sqrt{R^4+4G^2J^2}}
 \, .\end{equation}
This functional dependence of $k_r$ on $R$ and $J$ is exactly that
of the Kerr family. (The condition on surface gravity can always
be implemented provided the right side of (\ref{eq:kerrkappa}) is
positive, which in the kerr family corresponds to non-extremal
horizons. The resulting $r$ and $N_r$ are unique.) With this
choice of $N_r$ and $\Omega$, the energy $E^t_S$ is given by the
well known Smarr formula
\begin{equation} E^{t_o} = 2\, (\frac{k_o a}{8\pi G} + \Omega_o J) =
\frac{\sqrt{R^4 + 4G^2J^2}}{2GR} \, ;\end{equation}
as a function of its angular momentum and area, each cross-section
is assigned simply that mass which it would have in the Kerr
family. However, there is still a balance equation in which the
flux of gravitational energy $\mathcal{F}^{(t_o)}_g$ is local and
positive definite (see (\ref{eq:integrated1law})). (The
gravitational angular momentum flux which, in general, has
indeterminate sign vanishes due to axi-symmetry.) Motivated by the
isolated horizon framework, we will refer to this canonical
$E^{t_o}$ as the \textit{mass} associated with cross-sections $S$
of $H$ and denote it simply by $M$. Thus, among the infinitely
many first laws (\ref{eq:genfirstlaw}), there is a canonical one:
\begin{equation}  dM = \frac{k_o}{8\pi G} da + \Omega_o dJ \, .
\end{equation}

We conclude with three remarks.

i) Note that the mass and angular momentum depend only on
\emph{local} fields on each cross section $S$ and changes in these
quantities over \emph{finite} regions $\Delta H$ of $H$ have been
related to matter and gravitational radiation fluxes, determined
by the \emph{local} geometry of $H$.

ii) Unlike the vector fields $\xi^a_{(r)} = N_r \ell^a$, general
permissible vector fields $t^a$ is not necessarily causal.
Therefore the matter flux $\int_{\Delta H} T_{ab}t^a \th^{\, b}
d^3 V$ need not be positive. Similarly, if $\fie^a$ is not a
Killing field of $q_{ab}$, the gravitational flux need not be
positive. Therefore, although the area $a$ always increase along
$\hat{r}^a$, $E^t$ can decrease. This is the analog of the Penrose
process in which `rotational energy' is extracted from the
dynamical horizon.

iii) While the infinitesimal version eq~ (\ref{eq:genfirstlaw}) of
the first law is conceptually more interesting, the finite balance
equation (\ref{eq:integrated1law}) is likely to be more directly
useful in the analysis of astrophysical situations. In particular,
the presence of an infinite number of these balance equations can
provide useful checks on numerical simulations in the strong field
regime.

\textbf{Acknowledgements} We would like to thank Chris Beetle,
Steve Fairhurst and Jerzy Lewandowski for stimulating discussions.
This work was supported in part by the NSF grant PHY-0090091, the
NSF Cooperative Agreement PHY-0114375 and the Eberly research
funds of Penn State. BK was also supported through Duncan and
Roberts fellowships.

\end{document}